\documentclass[aps,twocolumn]{revtex4}
\usepackage{amsmath,amssymb}

\begin{document}
\title{Deep Connection Between Thermodynamics and Gravity in Gauss-Bonnet Braneworld}

\author{Ahmad Sheykhi $^{1,2}$\footnote{
asheykhi@mail.uk.ac.ir}, Bin Wang $^{1}$\footnote{
wangb@fudan.edu.cn}, Rong-Gen Cai
$^{3}$\footnote{cairg@itp.ac.cn}}
\address{$^1$  Department of Physics, Fudan University, Shanghai 200433, China\\
         $^2$  Physics Department and Biruni Observatory, Shiraz University, Shiraz 71454,
         Iran\\
$^{3}$ Institute of Theoretical Physics, Chinese Academic of
Sciences, P.O. Box 2735, Beijing 100080, China}
\begin{abstract}
We disclose the deep connection between the thermodynamics and
gravity in a general braneworld model with curvature correction
terms on the brane and in the bulk, respectively. We show that the
Friedmann equations on the $3$-brane embedded in the $5$D
spacetime with curvature correction terms can be written directly
in the form of the first law of thermodynamics on the apparent
horizon. Using the first law, we extract the entropy expression of
the apparent horizon on the brane, which is useful in studying the
thermodynamical properties of the black hole horizon on the brane
in Gauss-Bonnet gravity.
\end{abstract}

 \maketitle
\newpage

Inspired by the black hole physics that the first law of black
hole thermodynamics indicating some deep relations between
thermodynamics and gravitational Einstein equations, a profound
connection between gravity and thermodynamics has been argued to
exist. A pioneer work on this respect was done by Jacobson who
disclosed that the gravitational Einstein equation can be derived
from the relation between the horizon area and entropy, together
with the Clausius relation $\delta Q=T\delta S$ \cite{Jac}. This
derivation suggests that a hyperbolic second order partial
differential Einstein equation for the spacetime metric has a
predisposition to thermodynamic behavior.

Is the obtained relation between thermodynamics and the Einstein
equation just an accident? Does the connection between
thermodynamics and gravity hold in all spacetimes and even beyond
Einstein gravity? Does it imply something in deep? There have been
a lot of effort to answer these questions including studies in the
so called $f(R)$ gravity\cite{Elin}, Gauss-Bonnet gravity, the
scalar-tensor gravity and more general Lovelock
gravity\cite{Cai1}\cite{Pad} etc. The study has also been
generalized to the cosmological context
\cite{Cai2,Cai3,CaiKim,Fro,Dan,verlinde}, where it has been shown
that the differential form of the Friedmann equation of the FRW
universe at the apparent horizon can be rewritten in the form of
the first law of thermodynamics. The extension of the connection
between thermodynamics and gravity has also been carried out in
the braneworld cosmology \cite{Cai4}\cite{SW}.

In this work we will address the question on the connection
between thermodynamics and gravity by investigating whether and
how the relation can be found in general braneworld models with
correction terms, such as a 4D scalar curvature from induced
gravity on the brane, and a 5D Gauss-Bonnet curvature term. With
these correction terms, especially including a Gauss-Bonnet
correction to the 5D action, we have the most general action with
second-order field equations in 5D \cite{lovelock}, which provides
the most general models for the braneworld scenarios
\cite{RS}\cite{DGP}. Furthermore, in an effective action approach
to the string theory, the Gauss-Bonnet term corresponds to the
leading order quantum corrections to gravity, and its presence
guarantees a ghost-free action\cite{zwiebach}. The purpose of this
work is to disclose the connection between thermodynamics and
gravity in this very general braneworld scenario and show that
this connection is not an accident result in some specific
spacetime.

The motivating idea for disclosing the connection between the
thermodynamics and gravity is to get deeper understanding on the
entropy of the Gauss-Bonnet braneworld. In the braneworld
scenarios, gravity on the brane does not obey Einstein theory,
thus the usual area formula for the black hole entropy does not
hold on the brane. The exact analytic black hole solutions on the
brane have not been found so far, so that the relation between the
braneworld black hole horizon entropy and its geometry is not
known. We expect that the connection between gravity and
thermodynamics in the braneworld can shed some lights on
understanding these problems. This serves as our main motivation
to explore the thermodynamical properties of the Friedmann
equation in the braneworld with curvature correction terms.

We start from the following action
\begin{eqnarray}\label{Act}
S  &=&  \frac{1}{2{\kappa_5}^2} \int{
d^5x\sqrt{-{g}}\left({R}-2\Lambda+\alpha
\mathcal{L}_{GB}\right)} \nonumber\\
 && + \frac{1}{2{\kappa_4}^2}\int
{d^4x\sqrt{-\widetilde{g}}\widetilde{R}}+\int
{d^{4}x\sqrt{-\widetilde{g}}( \mathcal {L}_{m} -2 \lambda)},
\nonumber
\end{eqnarray}
where $\Lambda<0$ is the bulk cosmological constant and ${\mathcal
L}_{GB}=R^2-4R^{AB}R_{AB}+R^{ABCD}R_{ABCD}$ is the Gauss-Bonnet
correction term. Objects corresponding to the brane are written
with a tilde to be distinguished from $5$D objects; $g$ is the
bulk metric and $R$, $R_{AB}$, and $R_{ABCD}$ are the curvature
scalar, Ricci and Riemann tensors respectively. $\kappa_4$ and
$\kappa_5 $ are the gravitational constants on the brane and in
the bulk, respectively. The last term in (\ref{Act}) corresponds
to the matter content. $\mathcal{L}_m$ is the Lagrangian density
of the brane matter fields, and $\lambda$ is the brane tension (or
the brane cosmological constant). Hereafter we assume that the
brane cosmological constant is zero (if it does not vanish, one
can absorb it to the stress-energy tensor of matter fields on the
brane). We assume that there are no sources in the bulk other than
$\Lambda$ and redefine $\kappa_{4}^2=8\pi G_{4}\,
,\quad\kappa_{5}^2=8\pi G_{5}$. The field equations can be
obtained by varying action (\ref{Act}) with respect to the bulk
metric $g_{AB}$. The result is
\begin{equation}
  G_{AB}+\Lambda g_{AB}+2\alpha
H_{AB} = {\kappa_5}^2 T_{AB}\delta(y), \nonumber
 \end{equation}
 where we have
assumed that the brane is located at $y=0$ with $y$ being the
coordinate for the extra dimension, and
$H_{AB}=RR_{AB}-2R_A{}^CR_{BC}-2R^{CD}R_{ACBD}+R_A{}^{CDE}R_{BCDE}-\textstyle{1\over4}g_{AB}{\cal
L}_{GB} $ is the second-order Lovelock tensor. The energy-momentum
tensor can be further decomposed into two parts ${T}_{AB}=
\widetilde{T}_{AB}-\frac{1}{{\kappa_4}^2}
 \widetilde{G}_{AB}$,
where $\widetilde{T}_{AB}$ is the energy-momentum tensor
describing the matter confined on the 3-brane which we assume in
the form of a perfect fluid for homogenous and isotropic universe
on the brane $ \widetilde{T}_{AB} = (\rho+p)u_A u_B +
p\widetilde{g}_{AB} $, with $u^A$, $\rho$, and $p$, being the
fluid velocity ($u^A u_A=-1$), energy density and pressure
respectively. The possible contribution of nonzero brane tension
$\lambda$ will be assumed to be included in $\rho$ and $p$. A
homogeneous and isotropic brane at fixed coordinate position $y=0$
in the bulk is given by the five-dimensional line element
\begin{equation}
ds^2=-N^2(t,y) dt^2 + A^2(t,y)\gamma_{ij}dx^i dx^j
 + B^2(t,y)dy^2,  \nonumber
 \end{equation}
where $\gamma _{ij}$ is a maximally symmetric $3$-dimensional
metric for the surface ($t$=const., $y$=const.), whose spatial
curvature is parameterized by k = -1, 0, 1. On every hypersurface
($y$=const), we have the metric of a FRW cosmological model. The
metric coefficient $N$ is chosen so that, $N(t,0)=1$ and $t$ is
the cosmic time on the brane. The presence of the four-dimensional
curvature scalar in the gravitational action does not affect the
bulk equations. If we define $
\Phi=\frac{1}{N^{2}}\frac{\dot{A}^{2}}{A^{2}}-\frac{1}{B^{2}}
\frac{A^{\prime \,2}}{A^{2}}+\frac{k}{A^{2}} $ then the field
equation reduces to \cite{germani}
\begin{eqnarray}\label{Field1}
&& \frac{\dot{A}}{A}\frac{N'}{N}+\frac{A'}{A}\frac{\dot{B}}{B}-
\frac{\dot{A}'}{A}=0,  \nonumber \\
&&\Phi+2\alpha\Phi^{2}+\frac{1}{\ell^2}-\frac{\mathcal{C}}{A^{4}}=0,\label{Field2}
\end{eqnarray}
where $\ell^2\equiv-6/\Lambda$ and $\mathcal{C}$ is an integration
constant, which is related to the mass of the bulk black hole. In
the above equations dot and prime denote derivatives with respect
to $t$ and $y$, respectively. Integrating the $(00)$ component of
the field equation across the brane and imposing $\mathbb{Z}_2$
symmetry, we have the jump across the brane \cite{kofin}
\begin{eqnarray}\label{Jc1}
&&\frac{2{\kappa_4}^2}{{\kappa_5}^2}\left[
1+4\alpha\left(H^2+\frac{k}{a^{2} }- \frac{A^{\prime
\,2}_{+}}{3a^2b^2 }
\right)\right]\frac{A'_{+}}{ ab } \nonumber \\
 &&=-\frac{\kappa^{2}_{4}}{3}\rho+H^2 +\frac{k}{a^{2}},
\end{eqnarray}
where $a=A(t,0) $, $b=B(t,0) $ and $ 2A'_{+}=-2A'_{-}$ is the
discontinuity of the first derivative. $H=\dot a/a$ is the Hubble
parameter on the brane. Inserting $\Phi$ into Eq. (\ref{Jc1}) we
can obtain the generalized Friedmann equation on the
brane~\cite{kofin}
 \begin{eqnarray}\label{GFri}
&&\epsilon\frac{2{\kappa_4}^2}{{\kappa_5}^2}\left[1
+\frac{8}{3}\alpha\left(H^2 +{k \over a^2} + {\Phi_{0}\over 2}
\right) \right]\left(H^2 +{k \over a^2}-\Phi_{0}\right)^{1/2} \nonumber \\
&&=-\frac{\kappa^{2}_{4}} {3}\rho+H^2 +{k \over a^2} ,
\label{3fried}
 \end{eqnarray}
where $\Phi_0=\Phi(t,0)$ and $\epsilon=\pm1$. For later
convenience we choose $\epsilon=-1$.  This Friedmann equation
contains various special cases discussed extensively in the
literature. The DGP braneworld is the limiting case when
$\alpha=0$, while the RS II braneworld can be reproduced in the
limit $\kappa_{4}\to \infty$ and $\alpha=0$. The pure Gauss-Bonnet
braneworld is the case with $\kappa_4 \to \infty$. The Friedmann
equation~(\ref{GFri}) together with the energy-momentum tensor
conservation law $ \dot{\rho}+3(\rho+p)H=0, $ and the equation of
state $p=p(\rho)$, describe completely  the cosmological dynamics
on the brane. We will use the continuity and the Friedmann
equations to obtain the first law of thermodynamics at the
apparent horizon on the brane.

To have further understanding about the nature of the apparent
horizon we rewrite more explicitly, the metric of homogenous and
isotropic FRW universe on the brane in the form $ ds^2={h}_{\mu
\nu}dx^{\mu} dx^{\nu}+\tilde{r}^2d{\Omega_{2}}^2, $ where
$\tilde{r}=a(t)r$, $x^0=t, x^1=r$, the two dimensional metric
$h_{\mu \nu}$=diag $(-1, a^2(t)/(1-kr^2))$. Then, the dynamical
apparent horizon, is determined by the relation $h^{\mu
\nu}\partial_{\mu}\tilde {r}\partial_{\nu}\tilde {r}=0$, which
implies that the vector $\nabla \tilde {r}$ is null on the
apparent horizon surface. The apparent horizon is argued to be a
causal horizon for a dynamical spacetime and is associated with
gravitational entropy and surface gravity \cite{Hay2,Bak}. A
simple calculation gives the apparent horizon radius for the FRW
universe $
 \tilde{r}_A=\frac{1}{\sqrt{H^2+k/a^2}}.
$
The associated temperature with the apparent horizon can be defined
as $T = \kappa/2\pi$, where $\kappa$ is the surface gravity
$
 \kappa =\frac{1}{\sqrt{-h}}\partial_{\mu}\left(\sqrt{-h}h^{\mu \nu}\partial_{\mu\nu}\tilde
 {r}\right).
$ Then one can easily show that the surface gravity at the
apparent horizon of FRW universe can be written as $
\kappa=-\frac{1}{\tilde r_A}\left(1-\frac{\dot {\tilde
r}_A}{2H\tilde r_A}\right)$. When $\dot {\tilde r}_A\leq 2H\tilde
r_A$, the surface gravity $\kappa\leq 0$, which leads the
temperature $T\leq 0$ if one defines the temperature of the
apparent horizon as $T=\kappa/2\pi$ . Physically it is not easy to
accept the negative temperature, the temperature on the apparent
horizon should be defined as $T=|\kappa|/2\pi$.

In the remaining part of this paper we will apply these results to
investigate the thermodynamical properties of apparent horizon in
the Gauss-Bonnet braneworld scenario. Indeed we want to show that
the first law of thermodynamics on the apparent horizon can be
extracted directly from the Friedmann equations and therefore we
can extract an entropy expression associated with the horizon
geometry on the brane. To do this, first we assume there is no
black hole in the bulk and so $\mathcal{C}=0$. Inserting this
condition into Eq. (\ref{Field2}), we get $
 \Phi=\Phi_0=\frac{1}{4\alpha}\left(-1+\sqrt{1-\frac{8\alpha}{\ell^2}}\right)=\rm {const}.
$ This condition also implies $ \alpha<\ell^2/8$.  In terms of the
apparent horizon radius, we can rewrite the Friedmann equation
(\ref{GFri}) as
\begin{eqnarray}
\label{Fri1}
 \rho&=&\frac{3}{8\pi G_4}\frac{1}{{\tilde {r}_A}^2}+
\frac{3}{4\pi G_5} \left(\frac{1}{{\tilde {r}_{A}}^2}
-\Phi_{0}\right)^{1/2} \nonumber \\
&&\times\left[1 +\frac{8}{3}\alpha\left(\frac{1}{{\tilde
{r}_{A}}^2} + {\Phi_{0}\over 2} \right) \right].
 \end{eqnarray}
 Taking differential form of
equation (\ref{Fri1}) and using the continuity equation, we can
get the differential form of the Friedmann equation
\begin{eqnarray} \label{Fri2}
 H(\rho+p)dt=\frac{1}{4\pi G_{4}} \frac{d\tilde {r}_{A}}{\tilde
{r}_{A}^3}+\frac{1}{4\pi G_{5}}\frac{d\tilde {r}_{A}}{\tilde
{r}_{A}^2\sqrt{1-\Phi_0 \tilde {r}_{A}^2}} \nonumber \\
+\frac{1}{4\pi G_{5}}\frac{4\alpha}{\tilde
{r}_{A}^4}\left(\frac{2-\Phi_0 \tilde {r}_{A}^2 }{\sqrt{1-\Phi_0
\tilde {r}_{A}^2}}\right)d\tilde {r}_{A}.
\end{eqnarray}
Multiplying both sides of the equation (\ref{Fri2}) by a factor
$3\Omega_{3}\tilde{r}_{A}^{3}\left(1-\frac{\dot {\tilde r}_A}{2H
\tilde r_A}\right)$, and using the expression for the surface
gravity, after some simplifications we can rewrite this equation
in the form
\begin{eqnarray}\label{Fri3}
&&3H(\rho+p)\Omega_{3}\tilde{r}_{A}^{3}dt-\frac{3}{2}(\rho+p)\Omega_{3}\tilde{r}_{A}^{2}dt = \nonumber \\
&&~~~-\frac{\kappa}{2\pi}\left(\frac{3\Omega_{3}}{2G_{4}}\tilde{r}_{A}d\tilde{r}_{A}
+\frac{3\Omega_{3}}{2G_{5}}\frac{\tilde{r}_{A}^{2}d\tilde
{r}_{A}}{\sqrt{1-\Phi_0 \tilde {r}_{A}^2}}\right) \nonumber \\
&&-\frac{\kappa}{2\pi}\frac{6\alpha\Omega_{3}}{G_{5}}\left(\frac{2-\Phi_0
\tilde {r}_{A}^2 }{\sqrt{1-\Phi_0 \tilde {r}_{A}^2}}\right)d\tilde
{r}_{A}.
 \end{eqnarray}
Next, we take $E=\rho V$ the total energy of the matter on the
brane inside a 3-ball with the volume
$V=\Omega_3\tilde{r}_{A}^{3}$. The area of the apparent horizon is
the area of the ball on a 2-sphere of radius $\tilde{r}_{A}$,
which is expressed as $A=3\Omega_3\tilde{r}_{A}^2$\cite{TT}.
Taking differential form of the relation $ E=\rho
\Omega_{3}\tilde{r}_{A}^{3}$ for the total matter energy inside
the apparent horizon on the brane, we get $
 dE = 3\Omega_{3}\tilde
 {r}_{A}^{2}\rho d\tilde {r}_{A}+\Omega_{3}\tilde
 {r}_{A}^{3}\dot {\rho} dt.
$ Using the continuity relation, we obtain $
 dE=3\Omega_{3}\tilde
 {r}_{A}^{2}\rho d\tilde {r}_{A}-3 H(\rho+p)\Omega_{3}\tilde
 {r}_{A}^{3}dt.
$
Substituting this relation into (\ref{Fri3}), and using the
relation between temperature and the surface gravity, we can
rewrite this equation in the form
\begin{eqnarray}\label{Fri4}
dE -WdV &=& T \left(\frac{3\Omega_{3}}{2G_{4}}\tilde{r}_{A}
+\frac{3\Omega_{3}}{2G_{5}}\frac{\tilde{r}_{A}^{2}}{\sqrt{1-\Phi_0 \tilde {r}_{A}^2}}\right)d\tilde {r}_{A} \nonumber \\
& + & T \ \frac{6\alpha\Omega_{3}}{G_{5}}\left(\frac{2-\Phi_0
\tilde {r}_{A}^2 }{\sqrt{1-\Phi_0 \tilde {r}_{A}^2}}\right)d\tilde
{r}_{A},
\end{eqnarray}
where $W=(\rho-p)/2$ is the matter work density defined by
$W=-\frac{1}{2} \widetilde{T}^{\mu \nu}h_{\mu \nu}$ \cite{Hay2}.
The work density term is regarded as the work done by the change
of the apparent horizon, which is used to replace the negative
pressure if compared with the standard first law of
thermodynamics, $dE = TdS - PdV$. For a pure de Sitter space,
$\rho=-p$, then our work term reduces to the standard $pdV$. Eq.
(\ref{Fri4}) is nothing, but the first law of thermodynamics at
the apparent horizon on the brane. That is, it can be rewritten as
\begin{equation}
dE=TdS+WdV,
\end{equation}
if we identify the entropy associated with the apparent horizon on
the brane as
\begin{eqnarray}\label{ent1}
 S &=& \frac{3\Omega_{3}}{2G_{4}}{\displaystyle\int^{\tilde
r_A}_0\tilde{r}_{A}d\tilde {r}_{A}}+
\frac{3\Omega_{3}}{2G_{5}}{\displaystyle\int^{\tilde r_A}_0
\frac{\tilde{r}_{A}^{2}d\tilde {r}_{A}}{\sqrt{1-\Phi_0 \tilde
{r}_{A}^2}}}\nonumber \\
&+& \frac{6\alpha\Omega_{3}}{G_{5}}{\displaystyle\int^{\tilde
r_A}_0 \frac{2-\Phi_0 \tilde {r}_{A}^2 }{\sqrt{1-\Phi_0 \tilde
{r}_{A}^2}}d\tilde {r}_{A}}.
\end{eqnarray}
The explicit form of the entropy can be obtained by integrating
(\ref{ent1}). The result is
\begin{eqnarray} \label{ent2}
S
&=&\frac{3\Omega_{3}{\tilde{r}_A}^{2}}{4G_{4}}+\frac{2\Omega_{3}{\tilde{r}_A}^{3}}{4G_{5}}
\times {}_2F_1\left(\frac{3}{2},\frac{1}{2},\frac{5}{2},
\Phi_0{\tilde{r}_A}^2\right) \nonumber \\
 && +\frac{6\alpha\Omega_{3}{\tilde{r}_A}^3}{G_{5}} \left(
\Phi_0 \times {}_2F_1\left(\frac{3}{2},\frac{1}{2},\frac{5}{2},
\Phi_0{\tilde{r}_A}^2\right)   \right. \nonumber
\\
&& \left.+\frac{\sqrt{1-\Phi_0 \tilde
{r}_{A}^2}}{{\tilde{r}_A}^2}\right),
\end{eqnarray}
where ${}_2F_1(a,b,c,z)$ is a hypergeometric function. The
expression looks complicated. But its physical meaning is clear.
Some remarks are as follows.

i) The first term in (\ref{ent2})has the form of
Bekenstein-Hawking entropy of the 4D Einstein gravity. In fact,
the Einstein-Hilbert term ($\tilde R$) on the brane contributes to
this area term. The second term is due to the Einstein-Hilbert
term ($R$) in the bulk and obeys the $5$-dimensional area formula
in the bulk \cite{Cai4,SW}. The Gauss-Bonnet correction term in
the bulk contributes to the last term.

ii) In the limit $\alpha\rightarrow 0$ $(\Phi_0=-\ell^{-2})$, as
expected, this expression reduces to the entropy of apparent
horizon in warped DGP braneworld embedded in a $AdS_5$ bulk
\cite{SW}
\begin{eqnarray}
S=\frac{3\Omega_{3}{\tilde{r}_A}^{2}}{4G_{4}}+\frac{2\Omega_{3}{\tilde{r}_A}^{3}}{4G_{5}}
 \times
{}_2F_1\left(\frac{3}{2},\frac{1}{2},\frac{5}{2},
-\frac{{\tilde{r}_A}^2}{\ell^2}\right).
\end{eqnarray}

iii) While in the limit $\alpha \rightarrow 0$ and $\Phi_0
\rightarrow 0$ (or equality $\ell \rightarrow \infty$), it reduces
to the entropy of apparent horizon in pure DGP braneworld with a
Minkowskian bulk
\begin{eqnarray}
 S
=\frac{3\Omega_{3}{\tilde{r}_A}^{2}}{4G_{4}}+\frac{2\Omega_{3}{\tilde{r}_A}^{3}}{4G_{5}}.
\end{eqnarray}
Clearly, the first and second terms  obey the area formula in $4$D
and $5$D, respectively. The second term comes due to the bulk
effect. In global, the apparent horizon on the brane can be
extended to the bulk.  In $\alpha\rightarrow 0$, the gravity in
the bulk is the 5D Einstein gravity so that one has the entropy
associated with the area of apparent horizon in the bulk [12].
Details of the extension of apparent horizon into the bulk and the
calculation of the apparent horizon area in the bulk was described
in [12]. The factor $2$ in the second term, comes from the
$\mathbb{Z}_2$ symmetry property of the bulk. The first term
corresponds to the gravity on the brane and the second term
corresponds to the gravity in the bulk. This indeed reflects the
fact that there are two gravity terms in the action of DGP model.

iv) Furthermore, taking the limit $\alpha \rightarrow 0$ and  $G_4
\rightarrow \infty $, while keeping $G_5$ finite, the first and
the last terms in (\ref{ent2}) vanish and we obtain the entropy
associated with the apparent horizon in RS II braneworld
\cite{Cai4,SW}
\begin{eqnarray}
S=\frac{2\Omega_{3}{\tilde{r}_A}^{3}}{4G_{5}}
 \times
{}_2F_1\left(\frac{3}{2},\frac{1}{2},\frac{5}{2},
-\frac{{\tilde{r}_A}^2}{\ell^2}\right).
\end{eqnarray}

v) Finally, keeping $\alpha$ finite, and taking the limit $G_4
\rightarrow \infty $ and  $\Phi_0 \rightarrow 0$, one can extract
from Eq. (\ref{ent2}) the entropy associated with the apparent
horizon on the brane in the Gauss-Bonnet braneworld with a
Minkowskian bulk
\begin{eqnarray}
 S=\frac{2\Omega_{3}{\tilde{r}_A}^{3}}{4G_{5}}
 \left(1+\frac{12 \alpha}{{\tilde{r}_A}^{2}}\right).
\end{eqnarray}
This gives an expression of horizon entropy in the Gauss-Bonnet
gravity~\cite{GB,Cai2,Cai3,CaiKim,Cvetic}. This is an expected
result. Indeed, because of the absence of scalar curvature term on
the brane and the negative cosmological constant in the bulk, no
localization of gravity happens on the brane. As a result, the
gravity on the brane is still $5$D Gauss-Bonnet gravity and the
brane looks like a domain wall moving in a Minkowski spacetime.
Therefore, the entropy of apparent horizon on the brane still
obeys the entropy formula in the bulk. Note that the factor $2$ in
the entropy comes from the $\mathbb{Z}_2$ symmetry of the bulk.


In summary, we have disclosed the deep connection between the
thermodynamics and gravity in the general braneworld model by
showing that the Friedmann equations on the $3$-brane embedded in
a $5$D spacetime with curvature correction terms, such as a $4$D
scalar curvature term on the brane and a $5$D Gauss-Bonnet term in
the bulk, can be written directly in the form of the first law of
thermodynamics, $dE=TdS+WdV$, at the apparent horizon. Noting that
here $E$ is not the Misner-Sharp energy, but the matter energy
$\rho V$ inside the apparent horizon and they are equal only in
Einstein gravity~\cite{Cai1,Cai2,Cai3}. The result obtained here
in the very general braneworld model further supports the idea
that gravitation on a macroscopic scale is a manifestation of
thermodynamics. The thermodynamic derivation of the gravitational
field equations in the very general braneworld model indicates
that the connection between thermodynamics and gravity is not just
an accident, but something with deep physical meaning. On the
other hand the disclosed relation on the Friedmann equation and
the first law of thermodynamics on the apparent horizon also sheds
the light on holography, since the Friedmann equation persists the
information in the bulk and the first law of thermodynamics on the
apparent horizon contains the information on the boundary. Our
derivation in the general braneworld model supports the argument
in \cite{verlinde}. Indeed, our study shows that the approach here
is powerful to find an expression of entropy related to the
horizon geometry. It could help to extract an expression of
entropy associated with the apparent horizon on the brane, which
is useful in studying the thermodynamical properties of black
holes on the brane in the general braneworld model with curvature
correction terms. We expect to confirm that the entropy
(\ref{ent2}) is just the expression of black hole horizon entropy
in this braneworld scenario.


\acknowledgments This work was partially supported by NNSF of
China, Ministry of Education of China and Shanghai Educational
Commission. RGC was supported partially by grants from NSFC, China
(No. 10325525 and No. 90403029), and a grant from the Chinese
Academy of Sciences.


\begin{thebibliography}{99}
\bibitem{TT} $\Omega_{3}=\frac{\pi^{3/2}}{\Gamma(5/2)}=4\pi/3$.
For the benefit to generalize our results to the $(n-1)$-brane
embedded in the $(n+1)$-dimensional spacetime and compare our
results with those in [11,12], we use $\Omega_3$ here.
\end{thebibliography}
\end{document}